\documentclass[12pt,a4paper]{article}

\usepackage{hyperref}
\usepackage[numbers,sort&compress]{natbib}
\usepackage{graphicx}
\usepackage{bbm}
\usepackage{amsmath,amssymb}
\usepackage{xspace}

\usepackage{microtype}

\makeatletter
\let\old@startsection=\@startsection
\let\oldl@section=\l@section
\renewcommand{\@startsection}[6]{\old@startsection{#1}{#2}{#3}{#4}{#5}{#6\mathversion{bold}}}
\renewcommand{\l@section}[2]{\oldl@section{\mathversion{bold}#1}{#2}}
\makeatother

\makeatletter
\let\old@makecaption=\@makecaption
\def\@makecaption{\small\old@makecaption}
\makeatother

\renewcommand{\thefootnote}{\arabic{footnote}}
\let\oldPhi=\Phi
\let\oldPsi=\Psi
\let\oldGamma=\Gamma
\let\oldDelta=\Delta
\let\oldSigma=\Sigma
\let\oldTheta=\Theta
\let\oldPi=\Pi
\let\oldUpsilon=\Upsilon
\renewcommand{\Phi}{\mathnormal{\oldPhi}}
\renewcommand{\Psi}{\mathnormal{\oldPsi}}
\renewcommand{\Gamma}{\mathnormal{\oldGamma}}
\renewcommand{\Sigma}{\mathnormal{\oldSigma}}
\renewcommand{\Delta}{\mathnormal{\oldDelta}}
\renewcommand{\Theta}{\mathnormal{\oldTheta}}
\renewcommand{\Pi}{\mathnormal{\oldPi}}
\renewcommand{\Upsilon}{\mathnormal{\oldUpsilon}}

\newcommand{\superN}{\mathcal{N}}

\newcommand{\res}{\mathop{\mathrm{res}}}

\renewcommand{\Re}{\mathop{\mathrm{Re}}}

\newcommand{\order}{\mathcal{O}}

\newcommand{\Integers}{\mathbbm{Z}}
\newcommand{\Reals}{\mathbbm{R}}
\newcommand{\Complex}{\mathbbm{C}}
\newcommand{\Sphere}{S}  
\newcommand{\AdS}{\mathrm{AdS}}
\newcommand{\CP}{\mathbbm{CP}}
\newcommand{\RP}{\mathbbm{RP}}
\newcommand{\CFT}[1]{$\textrm{CFT}_{#1}$}
\newcommand{\SYM}[1]{$\textrm{SYM}_{#1}$}

\newcommand{\rep}[1]{{\mathbf{#1}}}

\newcommand{\grp}[1]{\mathrm{#1}}

\newcommand{\grSU}{\grp{SU}}
\newcommand{\grSO}{\grp{SO}}

\newcommand{\grOSp}{\grp{OSp}}

\newcommand{\lrbrk}[1]{\left(#1\right)}

\newcommand{\Bigbrk}[1]{\Bigl(#1\Bigr)}

\newcommand{\lrsbrk}[1]{\left[#1\right]}

\makeatletter
\def\mr@ignsp#1 {\ifx\:#1\@empty\else #1\expandafter\mr@ignsp\fi}%
\newcommand{\multiref}[1]{\begingroup
\xdef\mr@no@sparg{\expandafter\mr@ignsp#1 \: }%
\def\mr@comma{}%
\@for\mr@refs:=\mr@no@sparg\do{\mr@comma\def\mr@comma{,}\ref{\mr@refs}}%
\endgroup}
\makeatother

\newcommand{\hypref}[2]{\ifx\href\asklfhas #2\else\href{#1}{#2}\fi}

\newcommand{\appref}[1]{App.~\multiref{#1}}

\renewcommand{\eqref}[1]{(\multiref{#1})}

\ifx\href\asklfhas\newcommand{\href}[2]{#2}\fi

\def\XXint#1#2#3{{\setbox0=\hbox{$#1{#2#3}{\int}$}
    \vcenter{\hbox{$#2#3$}}\kern-.5\wd0}}

\newcommand{\conj}[1]{\bar{#1}}

\newcommand{\Luscher}{L{\"u}scher\xspace}

\begin{document}

\newpage
\setcounter{page}{1}
\renewcommand{\thefootnote}{\arabic{footnote}}
\setcounter{footnote}{0}


\thispagestyle{empty}
\begin{flushright}\footnotesize
\texttt{UUITP-22/08} \vspace{0.8cm}
\end{flushright}

\renewcommand{\thefootnote}{\fnsymbol{footnote}}
\setcounter{footnote}{0}

\begin{center}
{\Large\textbf{\mathversion{bold}
Finite size giant magnons in the $\grSU(2)\times\grSU(2)$ sector of $\AdS_4\times\CP^3$
}\par}

\vspace{1.5cm}

\textrm{Tomasz {\L}ukowski$^1$ and Olof Ohlsson Sax$^2$} \vspace{8mm}

\textit{$^1$
Institute of Physics, Jagellonian University\\
ul. Reymonta 4, 30-059 Krak\'ow, Poland}\\
\texttt{tomaszlukowski@gmail.com} \vspace{3mm}

\textit{$^2$
Department of Physics and Astronomy, Uppsala University\\
SE-75108 Uppsala, Sweden}\\
\texttt{olof.ohlsson-sax@physics.uu.se} \vspace{3mm}


\par\vspace{1cm}

\textbf{Abstract} \vspace{5mm}

\begin{minipage}{13cm}
  We use the algebraic curve and \Luscher's $\mu$-term to calculate
  the leading order finite size corrections to the dispersion relation
  of giant magnons in the $\grSU(2) \times \grSU(2)$ sector of $\AdS_4
  \times \CP^3$. We consider a single magnon as well as one magnon in
  each $\grSU(2)$. In addition the algebraic curve computation is
  generalized to give the leading order correction for an arbitrary
  multi-magnon state in the $\grSU(2) \times \grSU(2)$ sector.
\end{minipage}

\end{center}

\vspace{0.5cm}

\newpage
\setcounter{page}{1}
\renewcommand{\thefootnote}{\arabic{footnote}}
\setcounter{footnote}{0}
\tableofcontents


\section{Introduction}

During the last decade, a large amount of work has been put into the
understanding of the duality between $\superN=4$ super Yang-Mills and
type IIB string theory on $\AdS_5 \times \Sphere^5$
\cite{Maldacena:1997re,Gubser:1998bc,Witten:1998qj}. An important
discovery was that the theories on both sides of this correspondence
are governed by integrable structures \cite{Minahan:2002ve,
  Beisert:2003tq, Beisert:2003yb, Mandal:2002fs, Bena:2003wd,
  Kazakov:2004qf}.

Motivated by the development of new superconformal world-volume
theories for multiple M2-branes
\cite{Bagger:2006sk,Bagger:2007jr,Gustavsson:2007vu,Bagger:2007vi},
\citeauthor*{Aharony:2008ug} recently proposed a new class of
superconformal field theories in 2+1 dimensions with $\superN=6$
supersymmetry, which are conjectured to describe $N$ interacting
M2-branes in a background of $\AdS_4 \times \Sphere^7 / \Integers_k$
\cite{Aharony:2008ug,Benna:2008zy}. These ABJM theories have $\grSU(N)
\times \grSU(N)$ gauge theory, with Chern-Simons terms at level $k$
for the gauge fields, and allows a 't Hooft limit where $k, N \to
\infty$ with the coupling $\lambda = N/k$ fixed. In the large $k$
limit, the membrane theory is compactified so that the dual theory is
given by type IIA string theory on an $\AdS_4 \times \CP^3$ background.

Part of the success in the studies of the $\AdS_5$/\SYM4 duality lies
in the identification of the fundamental excitations in the two
theories. In the weak coupling regime these are magnons propagating
along the gauge theory spin-chain \cite{Minahan:2002ve}. At large
coupling, magnons with finite momentum evolve into giant magnons
\cite{Hofman:2006xt}, describing localized solitonic excitations on
the world-sheet. The integrability of the theories was essential in
these calculations.

Remarkably, integrable structures seem to appear also in the new
$\AdS_4$/\CFT3. \citet{Minahan:2008hf} showed that the two-loop
dilation operator of the scalar $\grSU(4)$ sector of the Chern-Simons
theory is equivalent to an integrable Hamiltonian, and conjectured a
set of Bethe equations valid for the full two-loop theory (see also
\cite{Bak:2008cp}). At strong coupling, the type IIA action has been
formulated in terms of a super-coset sigma model
\cite{Arutyunov:2008if,Stefanski:2008ik}, and using the pure spinor
formalism \cite{Bonelli:2008us,Fre:2008qc}. Additionally an algebraic
curve has been constructed \cite{Gromov:2008bz}.  Both of these limits
are incorporated in the proposed all-loop generalization of the Bethe
equations~\cite{Gromov:2008qe}. These Bethe equations have also been
derived from the proposed exact S-matrix of the theory
\cite{Ahn:2008aa}.\footnote{%
  Recently a mismatch between the string theory and Bethe ansatz
  results for the one-loop correction to spinning strings. See
  \cite{McLoughlin:2008ms,Alday:2008ut,Krishnan:2008zs,Gromov:2008fy,McLoughlin:2008he}
  for discussions of this issue.}

The spin-chain of ABJM differs from that of $\superN=4$ SYM in that
the $\grSU(4)$ representations alternate between adjacent sites.%
\footnote{Another important difference is that the scalars in ABJM
  transform as a $\rep{4}$ or a $\conj{\rep{4}}$ under the $\grSU(4)$
  R-symmetry, while in $\superN=4$ SYM they transform as a $\rep{6}$.}
The spin-chain ground state preserves an $\grSU(2|2)$ subgroup of the
full $\grOSp(2,2|6)$ symmetry of the gauge theory. The fundamental
excitations fall into two $(2|2)$ multiplets
\cite{Gaiotto:2008cg,Ahn:2008aa}. In addition there are quasi-bound
states. The theory has an important closed $\grSU(2) \times \grSU(2)$
subsector, which includes one excitation from each fundamental
multiplet.

At strong coupling, the spin-chain ground state corresponds to a
point-like string spinning on a great circle of each sphere
\cite{Gaiotto:2008cg,Nishioka:2008gz,Chen:2008qq,Gromov:2008bz}.
World-sheet excitations above this ground state have been studied in
the plane wave limit
\cite{Nishioka:2008gz,Gaiotto:2008cg,Grignani:2008is}. Additionally,
two different kinds of giant magnons have been found. The first one is
in $\Reals \times \Sphere^2 \times \Sphere^2$, where the magnons live
on one or both of the spheres
\cite{Gaiotto:2008cg,Grignani:2008is,Lee:2008ui,Shenderovich:2008bs,Ryang:2008rc,Berenstein:2008dc}. The
other giant magnon solution is spinning on $\Reals \times \RP^2$
\cite{Gaiotto:2008cg,Shenderovich:2008bs}. In this paper, only the
first kind of magnons will be considered.

In recent years, one aspect of the $\AdS_5$/\SYM4 duality that has
attracted much interest is that of finite size corrections and
wrapping interactions. The gauge theory spectrum derived from the
Bethe equations is valid only for asymptotically large operators. For
finite size operators, corrections are expected to arise
\cite{Ambjorn:2005wa}. Recently the four loop corrections stemming
from wrapping interactions have been calculated directly from the
gauge theory
\cite{Fiamberti:2007rj,Fiamberti:2008sh,Velizhanin:2008jd}, as well as
using the thermodynamic Bethe ansatz (TBA) and the \Luscher formulae
\cite{Bajnok:2008bm}.

On the string theory side, finite size corrections to the giant magnon
dispersion relation have been studied using direct sigma model
calculations \cite{Arutyunov:2006gs,Astolfi:2007uz}, \Luscher formulae
\cite{Janik:2007wt,Hatsuda:2008na}, the algebraic curves
\cite{Minahan:2008re} and analogies with the sine-Gordon equation
\cite{Klose:2008rx}.

For the $\AdS_4 \times \CP^3$ theory, finite size effects in the
Penrose limit have been considered \cite{Astolfi:2008ji}, and the
finite size corrections to the giant magnon dispersion relation have
been calculated for the case of two $\grSU(2) \times \grSU(2)$ magnons
with equal momenta \cite{Grignani:2008te,Lee:2008ui,Ahn:2008hj}. In
this paper we will consider finite size corrections to more general
multi-magnon states in the $\grSU(2) \times \grSU(2)$ sector. The
calculation of finite size effects using different formulations of the
theory pose a good consistency check.

\bigskip\noindent
While this paper was being prepared, we received
\cite{Bombardelli:2008qd} which contains results that overlap with
parts of this paper.

\section{Finite size corrections from the algebraic curve}

The algebraic curve for giant magnons in $\AdS_5\times\Sphere^5$ was
first given in \cite{Minahan:2006bd}, and was discussed in more detail
in \cite{Vicedo:2007rp}. In \cite{Minahan:2008re}, the curve for a
finite size magnon was constructed. Finite size corrections were also
discussed in a finite gap context in
\cite{Gromov:2008ec,Sax:2008in}. In this section we build upon
these solutions to obtain the energy shift for finite size giant
magnons in the $\grSU(2)\times\grSU(2)$ Chern-Simons theory.

\subsection{The algebraic curve}

Using the algebraic curve of \cite{Gromov:2008bz}, a classical string
state in $\AdS_4\times\CP^3$ is mapped to a ten-sheeted Riemann
surface. The branches $q_i(x)$, $i = 1, \dotsc, 10$ of this surface
are called the \textit{quasi-momenta} and are parametrized by a
spectral parameter $x \in \Complex$. Pairs of these sheets can be
connected by square root cuts $\mathcal C_{ij}$. When going through
the cut the quasi-momenta get shifted by an integer multiple
of $2\pi$
\begin{equation}
  q_i(x + i\epsilon) - q_j(x - i\epsilon) = 2\pi n_{ij},
\end{equation}
where $q_i$ and $q_j$ are evaluated on opposing side of the cut, and
$n_{ij} \in \Integers$ are called \textit{mode numbers}.

The charges of the string state corresponding to a specific curve is
given by the inversion symmetry and the curve's asymptotic behavior at
large $x$. Some important properties of the algebraic curve are summarized
in \appref{ap:prop-of-alg-curve}.

\subsection{Ansatz for the algebraic curve in the $\grSU(4)$ sector}

Our aim is to find quasi-momenta $q_{1}(x),\dotsc,q_{10}(x)$ with
the correct poles and symmetries, and having the right large $x$
asymptotics. In this paper we will treat the $\grSU(2)\times\grSU(2)
\subset \grSU(4)$ sector and use the  ansatz \cite{Gromov:2008qe}
\begin{align}
  q_1(x) = -q_{10}(x) &= \alpha \frac{x}{x^2 - 1}, \label{eq:SO6-ansatz-q1}\\
  q_2(x) = -q_9(x) &= \alpha \frac{x}{x^2 - 1}, \label{eq:SO6-ansatz-q2}\\
  q_3(x) = -q_8(x) &= \alpha \frac{x}{x^2 - 1} + G_r(x) + G_r\lrbrk{\frac{1}{x}} - G_v\lrbrk{\frac{1}{x}} - G_u\lrbrk{\frac{1}{x}} 
    \nonumber\\ &\qquad\qquad
    - G_r(0) + G_v(0) + G_u(0), \label{eq:SO6-ansatz-q3}\\
  q_4(x) = -q_7(x) &= \alpha \frac{x}{x^2 - 1} +G_v(x) + G_u(x) - G_r(x) - G_r\lrbrk{\frac{1}{x}} + G_r(0), \label{eq:SO6-ansatz-q4}\\
  q_5(x) = -q_6(x) &= -G_v(x) + G_u(x) - G_v\lrbrk{\frac{1}{x}} + G_u\lrbrk{\frac{1}{x}} + G_v(0) - G_u(0). \label{eq:SO6-ansatz-q5}
\end{align}
The subscripts of the resolvents $G_v$, $G_u$ and $G_r$ correspond to
the excitation numbers of \appref{sec:notation}, and indicate which
Dynkin labels of $\grSU(4)$ are excited by a cut in the resolvent.

\subsection{$\grSU(2)$ giant magnon}

As a simple check of the ansatz
(\ref{eq:SO6-ansatz-q1})-(\ref{eq:SO6-ansatz-q5}) we will derive the
dispersion relation of a single $\grSU(2)$ giant magnon. The
resolvents then take the form
\begin{align}
  G_v(x) &= \frac{1}{i} \log \frac{x - X^+}{x - X^-}, &
  G_u(x) &= G_r(x) = 0.
\end{align}
In order to obtain conserved charges of the magnon we have to consider
the large $x$ behavior of the quasi-momenta, and compare it with the
expected limits from \appref{ap:prop-of-alg-curve}%
\footnote{The coupling $g$ is related to the 't Hooft coupling $\lambda$ by 
  \begin{equation*}
    \lambda = 8 g^2.
  \end{equation*}}
\begin{align}
  q_{1}(x) = q_{2}(x) &= \frac{\alpha x}{x^2} + \dotsb = \frac{E \pm S}{2gx} + \dotsb,
  \label{eq:gm-small-E-eq}\\
  q_4(x) + q_3(x) &= -\frac{i}{x} \lrbrk{X^+ - X^- - \frac{1}{X^+} + \frac{1}{X^-} + 2 i \alpha} + \dotsb 
    = -\frac{J}{2gx}+ \dotsb 
  \label{eq:gm-small-J-eq} \\
  q_5(x) = q_4(x) - q_3(x) &= -\frac{i}{x} \lrbrk{X^+ - X^- + \frac{1}{X^+} - \frac{1}{X^-}} + \dotsb 
    = -\frac{Q}{2gx}+ \dotsb.
  \label{eq:gm-small-Q-eq}
\end{align}
and we can find from (\ref{eq:gm-small-E-eq}) that $E = 2 g \alpha$
and $S = 0$.  To check the inversion symmetry we
calculate\footnote{When considering a single giant magnon we can relax
  the level matching condition so that $p \not\in \pi\Integers$.}
\begin{equation}
  \pi m = q_3(1/x) + q_4(x) = -i \log\frac{X^+}{X^-} \equiv p.
  \label{eq:gm-small-p-eq}
\end{equation}
Solving \eqref{eq:gm-small-Q-eq} together with the momentum equation
\eqref{eq:gm-small-p-eq} for $X^\pm$ we get
\begin{equation}
  X^\pm = \frac{\frac{Q}{2} + \sqrt{\frac{Q^2}{4} + 16g^2\sin\frac{p}{2}}}{4g\sin\frac{p}{2}} \, e^{\pm i\frac{p}{2}}.
\end{equation}
Plugging this into \eqref{eq:gm-small-J-eq} gives the dispersion
relation
\begin{equation}
  E - \frac{J}{2} 
  = \sqrt{\frac{Q^2}{4} + 16 g^2 \sin^2\frac{p}{2}} 
  = \sqrt{\frac{Q^2}{4} + 2 \lambda \sin^2\frac{p}{2}}.
\end{equation}
This dispersion relation for the $\grSU(2)$ magnon is the same as the
``small'' giant magnon dispersion relation considered by
\citet{Gaiotto:2008cg} and by \citet{Shenderovich:2008bs}.

\subsubsection{Finite size corrections to $\grSU(2)$ giant magnon}

Let us continue by computing the finite size correction to a single magnon in
the $\grSU(2)$ sector. Inspired by \cite{Minahan:2008re} we use the
resolvents%
\footnote{These resolvents was used in \cite{Sax:2008in} to
  calculate the finite size corrections to the giant magnon dispersion
  relation in $\superN=4$ SYM.}
\begin{align}
  G_v(x) &= G(x) = -2i \log \frac{\sqrt{x-X^+} + \sqrt{x-Y^+}}{\sqrt{x-X^-} + \sqrt{x-Y^-}}, &
  G_u(x) &= G_r(x) = 0.
\end{align}
The function $G(x)$ has a log cut between the points $X^+$ and $X^-$
and two square root cuts connecting $X^\pm$ and $Y^\pm$. In the limit
$Y^\pm \to X^\pm$, the resolvent $G(x) \to -i\log\frac{x-X^+}{x-X^-}$,
which gives the previous single magnon solution.

The momentum of the magnon can be found from the inversion symmetry
\begin{equation}
  p = q_3(1/x) + q_4(x) = -2i\log\frac{\sqrt{X^+} + \sqrt{Y^+}}{\sqrt{X^-} + \sqrt{Y^-}}.
\end{equation}
and the conserved charges from the large $x$ asymptotics
\begin{align}
  \frac{J}{2g} &\approx \frac{E}{g} + \frac{i}{2} \lrbrk{X^+ - X^- + Y^+ - Y^- - \frac{2}{\sqrt{X^+ Y^+}} + \frac{2}{\sqrt{X^- Y^-}}},
  \label{eq:gm-small-J-eq-fin} \\
  \frac{Q}{2g} &\approx -\frac{i}{2} \lrbrk{X^+ - X^- + Y^+ - Y^- + \frac{2}{\sqrt{X^+ Y^+}} - \frac{2}{\sqrt{X^- Y^-}}}.
  \label{eq:gm-small-Q-eq-fin}
\end{align}
To solve the equations (\ref{eq:gm-small-J-eq-fin}) and (\ref{eq:gm-small-Q-eq-fin}) we introduce
\begin{equation}
  i \delta e^{i\phi} = Y^+ - X^+,
\end{equation}
and solve the equations perturbatively in $\delta$ (for $g \gg 1$). The result is
\begin{equation}
  E - \frac{J}{2} = 4 g \sin\frac{p}{2} - g\frac{\delta^2}{4} \sin\frac{p}{2} \cos(p - 2\phi).
\end{equation}

In order to calculate $\delta$ and $\phi$ we need to use the condition
that the sheets $q_4$ and $q_5$ are connected by square root cuts. This reads
\begin{equation}
  q_4(x + i\epsilon) - q_5(x - i\epsilon) = 2\pi n, \quad x \in \mathcal C,
\end{equation}
where $\mathcal C$ is one of the cuts. Focusing on the upper cut we get the condition
\begin{equation}
  2\pi n = \frac{E}{2g} \frac{x}{x^2-1} + G(x + i\epsilon) + G(x - i\epsilon) + G(1/x) - G(0).
  \label{eq:cut-cond-G}
\end{equation}
The first part of the right hand side is the same as in the
$\superN=4$ case,  so we can incorporate the result from that case,
which is
\begin{equation}\label{eq:res-from-n=4}
  G(x + i\epsilon) + G(x - i\epsilon) = 
  -2i\log\frac{Y^+-X^+}{x-X^-} + 4i\log \lrbrk{1 + \sqrt\frac{x-Y^-}{x-X^-}}.
\end{equation}
We are interested in the leading order behavior as $Y^\pm \to X^\pm$
in the formula (\ref{eq:res-from-n=4}). Hence we can evaluate it at $x
= X^+$. We then get
\begin{align*}
  \frac{E}{2g} \frac{x}{x^2-1} + G(x + i\epsilon) + G(x - i\epsilon) 
  &\approx \frac{E}{2g} \frac{X^+}{{X^+}^2-1} + G(X^+ + i\epsilon) + G(X^+ - i\epsilon) + \order(\delta) \\
  &\approx \frac{E}{2g} \frac{X^+}{{X^+}^2-1} - 2i \log\frac{i e^{i\phi} \delta}{4(X^+ - X^-)} + \order(\delta) \\
  &\approx -i \frac{E}{4g\sin\frac{p}{2}} - 2i\log\frac{e^{i\phi} \delta}{8\sin\frac{p}{2}} + \order(\delta).
\end{align*}

The last two terms in \eqref{eq:cut-cond-G} do not appear in the
$\superN=4$ case and need to be treated a bit more carefully. They are given by
\begin{align*}
  G(1/X^+) - G(0) &= -i \log\frac{\frac{1}{X^+} - X^+}{\frac{1}{X^+} - X^-}
    +i \log\frac{X^+}{X^-} + \order(\delta) \\
  &= -i\log\lrbrk{\cos\frac{p}{2} + i \sin\frac{p}{2} \frac{\sqrt{\frac{Q^2}{4} + 16g^2\sin^2\frac{p}{2}}}{\frac{Q}{2}}} - \frac{p}{2} + \order(\delta) \\
  &\approx -i\log\frac{8 i g \sin^2\frac{p}{2}}{Q} - \frac{p}{2} + \order(\delta).
\end{align*}   

Collecting the terms we get the condition
\begin{equation}
  2\pi n = -i \frac{E}{4g\sin\frac{p}{2}} - 2i\log\frac{e^{i\phi} \delta}{8\sin\frac{p}{2}}
           -i\log\frac{8 i g \sin^2\frac{p}{2}}{Q} - \frac{p}{2} + \order(\delta),
\end{equation}
which gives
\begin{equation}
  \delta = \sqrt\frac{8 Q}{g} e^{-\frac{E}{8g\sin\frac{p}{2}}}, \qquad
  \phi   = \frac{p}{4} + n \pi \pm \frac{\pi}{4},
\end{equation}
where the sign of the last term depends on how we chose the branch of $\frac{1}{2}\log\,i$.
The finite size dispersion relation is now given by
\begin{equation}
  E - \frac{J} = 4 g \sin\frac{p}{2} \pm 2Q \sin\frac{p}{2} \sin\lrbrk{\frac{p}{2} - 2\pi n} e^{-\frac{E}{4g\sin\frac{p}{2}}}.
  \label{eq:disp-rel-corr-alg-c}
\end{equation}
The form of this correction is very different from the one in the
$\superN=4$ case, since the leading order correction is suppressed by
a factor $1/g$ in addition to the exponential suppression. Moreover
the $\superN=4$ corrections are independent of the charge $Q$ for $Q
\ll g$. In the present case, the leading corrections vanish if we let
$Q \to 0$.

To identify more easily the correction we can consider a physical
state consisting of $M$ magnons with momentum $p$ and charge $Q$. This
is described by shifting the resolvent $G(x) \to M \cdot G(x)$. The
correction is now given by
\begin{equation}
  E - \frac{J} = 4 M g \sin\frac{p}{2} \lrsbrk{1 \pm \frac{Q}{2g} \sin\lrbrk{\frac{p}{2} - \frac{2\pi n}{M}} e^{-\frac{E/M}{4g\sin\frac{p}{2}}}}.
  \label{eq:disp-rel-corr-alg-c-M}
\end{equation}
For a physical configuration $p = \frac{\pi m}{M}$ for some integer
$m$. For a fundamental magnon ($Q = 1$) we get
\begin{align}
  \delta\mathcal E &= 2 \sin^2 \frac{p}{2} e^{-\frac{E}{4g\sin\frac{p}{2}}}, & n &= 0 \\ 
  \delta\mathcal E &= 0, & n &= \frac{p}{4\pi}.
\end{align}

\subsection{$\grSU(2)\times\grSU(2)$ giant magnon}

We now want to consider giant magnons in the $\grSU(2) \times
\grSU(2)$ sector. The simplest configuration consists of one
fundamental magnon in each $\grSU(2)$ sector, with equal momenta $p$.
For this case can use the ansatz
\eqref{eq:SO6-ansatz-q1}--\eqref{eq:SO6-ansatz-q5} with
\begin{equation}
  G_u(x) = G_v(x) = G(x) = -2i \log \frac{\sqrt{x-X^+} + \sqrt{x-Y^+}}{\sqrt{x-X^-} + \sqrt{x-Y^-}}
\end{equation}
and $G_r(x) = 0$. Following the same procedure as in the $\grSU(2)$
case this gives
\begin{equation}
  E - J = 8 g \sin\frac{p}{2} - g\frac{\delta^2}{2} \sin\frac{p}{2} \cos(p - 2\phi).
\end{equation}

Again we need to consider the condition that the quasi-momenta should
have square root cuts. The two cuts are at the same position, but
connect different sheets. In order to write down the condition we
imagine separating them slightly, so that we can consider two points
on opposite sides of one of the cuts, but on the same side of the
other. Our condition is then
\begin{equation}
  2\pi n = q_4(x + i\epsilon) - q_5(x - i\epsilon) 
  = \frac{E}{2g} \frac{x}{x^2 - 1} + G(x+i\epsilon) + G(x-i\epsilon).
  \label{eq:cut-cond-su2-su2}
\end{equation}
Note that the terms of the kind $G(1/x) - G(0)$ exactly cancel between
the two magnons. Equation \eqref{eq:cut-cond-su2-su2} is identical to
the corresponding equation in $\superN=4$, and the solution is
\begin{equation}
  \delta = 8 \sin\frac{p}{2} e^{-\frac{E}{8g\sin\frac{p}{2}}}, \qquad
  \phi = -\pi - \pi n.
\end{equation}
Thus the finite size dispersion relation for this configuration is
\begin{equation}
  \mathcal E = E - J = 8g\sin\frac{p}{2} \lrsbrk{
    1 - 4 \sin^2\frac{p}{2} \cos (p - 2\pi n) e^{-\frac{E}{4g\sin\frac{p}{2}}}}.
\end{equation}
Again a simple generalization to $M$ equal magnons in each sector
leads to two natural choices for $n$:
\begin{align}
  \delta \mathcal E &= -32 g \sin^3\frac{p}{2} \cos p \; e^{-\frac{E}{4g\sin\frac{p}{2}}}, & n &= 0, \\
  \delta \mathcal E &= -32 g \sin^3\frac{p}{2} e^{-\frac{E}{4g\sin\frac{p}{2}}}, & n &= \frac{p}{2\pi}.
\end{align}

\subsubsection{General multi-magnon states}

Using the algebraic curve we can also calculate the finite size
corrections to a general multi-magnon state in the $\grSU(2) \times
\grSU(2)$ sector. Hence we consider a state consisting of $M$ magnons
in the $\grSU(2)_v$ sector and $\hat M$ magnons in the $\grSU(2)_u$
sector, having momenta $p_i$ and $\hat p_i$ respectively.

At infinite $J$, the dispersion relation will be given by
\begin{equation}
  \mathcal E_\infty = \sum^M \mathcal E_i + \sum^{\hat M} \hat{\mathcal E}_i, 
  \qquad \mathcal E_i = 4g\sin\frac{p_i}{2},
  \quad \hat{\mathcal E}_i = 4g\sin\frac{\hat p_i}{2}.
\end{equation}
At finite $J$ this will get corrections, and we will write
\begin{equation}
  \mathcal E = \sum_{i=1}^M \Bigbrk{\mathcal E_i + \delta \mathcal E_i}
             + \sum_{i=1}^{\hat M} \Bigbrk{\hat{\mathcal E}_i + \delta\hat{\mathcal E}_i}.
\end{equation}

As an ansatz for the algebraic curve, we use a generalization of the
previous one with
\begin{align}
  G_v(x) &= \sum_{i=1}^M G_i(x) 
          = \sum_{i=1}^M \lrbrk{-2i \log \frac{\sqrt{x-X_i^+} + \sqrt{x-Y_i^+}}{\sqrt{x-X_i^-} + \sqrt{x-Y_i^-}}}, \\
  G_u(x) &= \sum_{i=1}^M \hat G_i(x)
          = \sum_{i=1}^{\hat M} \lrbrk{-2i \log \frac{\sqrt{x-\hat X_i^+} + \sqrt{x-\hat Y_i^+}}{\sqrt{x-\hat X_i^-} + \sqrt{x-\hat Y_i^-}}}.
\end{align}
For definiteness let us consider the first magnon in
$\grSU(2)_v$. Following the previous procedure we get
\begin{equation}
  \delta\mathcal E_1 = - g\frac{\delta^2}{4} \sin\frac{p_1}{2} \cos(p_1 - 2\phi).
\end{equation}
Again we calculate $\delta$ and $\phi$ by requiring that
\begin{equation}
  q_4(x+i\epsilon) - q_5(x-i\epsilon) = 2\pi n.
\end{equation}
Writing this out we get for $x$ in $\mathcal C^+_1$, the cut
connecting the branch points $X^+_1$ and $Y^+_1$,
\begin{multline}
    2\pi n = \frac{E}{2g} \frac{x}{x^2-1} + G_1(x + i\epsilon) + G_1(x - i\epsilon) + G_1(1/x) - G_1(0) \\
    + \sum_{i=2}^M \Bigbrk{G_i(1/x) - G_i(0)} - \sum_{i=1}^{\hat M} \Bigbrk{\hat G_i(1/x) - \hat G_i(0)}.
\end{multline}
The first row of this equation is identical to the one in the
one-magnon case. The second row induces interactions between the
magnons. From our previous results we have
\begin{multline}
  \frac{E}{2g} \frac{x}{x^2-1} + G_1(x + i\epsilon) + G_1(x - i\epsilon) + G_1(1/x) - G_1(0)
  \approx \\
  -i \frac{E}{4g\sin\frac{p_1}{2}} - 2i\log\frac{e^{i\phi} \delta}{8\sin\frac{p_1}{2}}
  -i\log\frac{8 i g \sin^2\frac{p_1}{2}}{Q_1} - \frac{p_1}{2} + \order(\delta).
\end{multline}
Moreover
\begin{align*}
  G_i\lrbrk{\frac{1}{x}} - G_i(0) 
  &\approx G_i\lrbrk{\frac{1}{X_1^+}} - G_i(0) \\
  &\approx -i \log \frac{\frac{1}{X_1^+}-X_i^+}{\frac{1}{X_1^+}-X_i^-} + i \log \frac{X_i^+}{X_i^-} \\
  &\approx -i \log \frac{\sin\frac{p_1 + p_i}{4}}{\sin\frac{p_1 - p_i}{4}} - \frac{p_i}{2},
\end{align*}
and similarly for $\hat G_i$. Thus
\begin{multline}
  \sum_{i=2}^M \Bigbrk{G_i(1/x) - G_i(0)} - \sum_{i=1}^{\hat M} \Bigbrk{\hat G_i(1/x) - \hat G_i(0)} \approx \\
  -i \log \lrbrk{ \prod_{i=2}^{M} \frac{\sin\frac{p_1 + p_i}{4}}{\sin\frac{p_1 - p_i}{4}} }
  +i \log \lrbrk{ \prod_{i=1}^{\hat M} \frac{\sin\frac{p_1 + \hat p_i}{4}}{\sin\frac{p_1 - \hat p_i}{4}}}
  - \sum_{i=2}^M \frac{p_i}{2} + \sum_{i=1}^{\hat M} \frac{\hat p_i}{2}.
\end{multline}
Collecting these results we get
\begin{multline}
  \delta\mathcal E_1 = 2Q_1 \sin\frac{p_1}{2} \;
  \prod_{i=2}^{M} \frac{\sin^2\frac{p_1 - p_i}{4}}{\sin^2\frac{p_1 + p_i}{4}} \;
  \prod_{i=1}^{\hat M} \frac{\sin^2\frac{p_1 + \hat p_i}{4}}{\sin^2\frac{p_1 - \hat p_i}{4}} \\
  \times \sin\lrbrk{p_1 - \sum_{i=1}^M \frac{p_i}{2} + \sum_{i=1}^{\hat M} \frac{\hat p_i}{2} + 2\pi n} \;
  e^{-\frac{E}{4g\sin\frac{p_1}{2}}}.
\end{multline}
As in $\superN=4$, the contribution from the magnon interactions is
related to the magnon S-matrix \cite{Minahan:2008re}. Note that
magnons in the same sector contribute with a different sign than
magnons in the opposite sector.

\section{Finite size corrections from the \Luscher $\mu$-term}
The second approach to the finite size effects is based on the so
called \Luscher formulae obtained for the first time by
\citet{Luscher:1985dn} for a relativistic field theory on a
cylinder and derived in \cite{Ambjorn:2005wa} for general
dispersion relations. We will focus only on the $\mu$-term which is
given by \cite{Janik:2007wt}
\begin{equation}
  \delta\mathcal E_a^\mu = -i \lrbrk{1 - \frac{\mathcal E'(p)}{\mathcal E'(\tilde q_*)}} \, e^{iq_*} 
  \cdot \res_{q=\tilde q} \sum_b S^{ba}_{ba}(q_*,p).
\end{equation}
Many of the following results can be easy obtained from the
$\AdS_5\times \Sphere^5$ case.

\subsection{$\grSU(2)$ giant magnon}

We start from the computations for an $\grSU(2)$ giant magnon. The
dispersion relation of a fundamental giant magnon in
$\AdS_4\times\CP^3$ is given by
\begin{equation}
  \mathcal E_4 = E - \frac{J}{2} = \sqrt{\frac{1}{4} + 16 g^2 \sin^2\frac{p}{2}},
\end{equation}
while the corresponding relation for the $\AdS_5\times\Sphere^5$ case is
\begin{equation}
  \mathcal E_5 = E - J = \sqrt{1 + 16 g^2 \sin^2\frac{p}{2}}.
\end{equation}
Note that $2\mathcal E_4$ equals $\mathcal E_5$ if we shift $g \to 2g$ and
$E \to 2E$ in $\mathcal E_5$. Hence we can import kinematical results
from $\superN=4$ to $\superN=6$, provided we make this shift of the
energy and the coupling. 

The matrix part cannot be obtained so easily from the
$\AdS_5\times\Sphere^5$ case so we have give it some more
attention. As described in \cite{Ahn:2008aa}, there are two types of
fundamental excitations in $\superN=6$ superconformal Chern-Simons
theory. We will refer to these as excitations of type $A$ and
$B$. Correspondingly the S-matrix can be divided into two parts -- the
matrices $S^{AA}$ and $S^{BB}$ describing scattering of particles of
the same type, and the matrices $S^{AB}$ and $S^{BA}$ describing
scattering of particles of different types. We write these S-matrices
as
\begin{align}
  S^{AA}(p_1,p_2) = S^{BB}(p_1,p_2) &= S_0(p_1,p_2) \hat S(p_1,p_2), \\
  S^{AB}(p_1,p_2) = S^{BA}(p_1,p_2) &= \tilde S_0(p_1,p_2) \hat S(p_1,p_2),
\end{align}
where $\hat S$ is the $\grSU(2|2)$-invariant S-matrix of
\cite{Arutyunov:2006yd} with $g$ appropriately shifted as noted
above. The scalar factors $S_0$ and $\tilde S_0$ are given by
\begin{align}
  S_0(p_1,p_2) &= \frac{1 - \frac{1}{x_1^+ x_2^-}}{1 - \frac{1}{x_1^- x_2^+}} \sigma(p_1,p_2), \\
  \tilde S_0(p_1,p_2) &= \frac{x_1^- - x_2^+}{x_1^+ - x_2^-} \sigma(p_1,p_2),
\end{align}
where $\sigma(p_1,p_2)$ is the BES dressing factor \cite{Beisert:2006ez}.

The relevant S-matrix coefficients are
\begin{align}
  a_1 &= \frac{x_2^- - x_1^+}{x_2^+ - x_1^-} \frac{\eta_1 \eta_2}{\tilde\eta_1 \tilde\eta_2} \\
  a_2 &= \frac{x_2^- - x_1^+}{x_2^+ - x_1^-} \frac{(x_1^- - x_1^+)(x_2^- - x_2^+)}{x_1^+x_2^+ - x_1^-x_2^-} \frac{\eta_1 \eta_2}{\tilde\eta_1 \tilde\eta_2} \\
  a_6 &= \frac{x_2^- - x_1^+}{x_2^+ - x_1^-} \frac{\eta_2}{\tilde\eta_2}.
\end{align}
The phase factors $\eta$ depend on the choice of basis. In the \textit{string frame}
\begin{equation}
  \frac{\eta_1}{\tilde\eta_1} = \sqrt{\frac{x_2^+}{x_2^-}}, \qquad
  \frac{\eta_2}{\tilde\eta_2} = \sqrt{\frac{x_1^-}{x_1^+}},
\end{equation}
while in the \textit{spin chain frame}
\begin{equation}
    \frac{\eta_1}{\tilde\eta_1} = \frac{\eta_2}{\tilde\eta_2} = 1.
\end{equation}

We will consider a single fundamental magnon of $A$-type. In order to
calculate the \Luscher $\mu$-term, we need to know the poles of the
S-matrix. Using the above expressions for the $\grSU(2)$ sector we see
that $S^{BA}(p_1,p_2)$ has no poles while $S^{AA}(p_1,p_2)$ has a
physical pole at $x_1^- = x_2^+$. The position of this pole is the
same as for a single $\grSU(2)$ magnon in $\superN=4$. Since the pole
positions agree, we can directly import the result for the kinematical
part from \cite{Janik:2007wt}. Thus
\begin{equation}
  \delta\mathcal E_a^\mu = -\frac{i}{2} \sin^2 \frac{p}{2} e^{-\frac{J}{8g\sin\frac{p}{2}}}
  \cdot \res_{q=\tilde q} \sum_b S^{ba}_{ba}(q_*,p).
\end{equation}

Following \cite{Janik:2007wt} we can express the S-matrix in terms of $a_i$
\begin{equation}
  \sum_b S^{ab}_{ab}(q_*,p) = S_0(q_*, p) (2a_1 + a_2 + 2a_6).
\end{equation}
and using the formulae for $a_i$ obtain the result which depends only on the frame we choose
\begin{align}
  \res_{q\to\tilde q} \sum_b S^{ab}_{ab}(q_*,p)
  &= \frac{1}{{x_1^-}'} \,\cdot\, \res_{x_1^- \to x_2^+} \sum_b S^{ab}_{ab}(q_*,p) \\
  &= \frac{i e^{-i\frac{p}{2}}}{\sin^2\frac{p}{2}} \,\cdot\, \res_{x_1^- \to x_2^+} \sum_b S^{ab}_{ab}(q_*,p) \\
  &= \frac{i}{g\sin^3\frac{p}{2}} \cdot \frac{\eta_1}{\tilde\eta_1}\frac{\eta_2}{\tilde\eta_2} \cdot \sigma(x_1,x_2).
\end{align}
Now we can plug it into the formula for $\mu$-term 
\begin{equation}
  \delta\mathcal E_a^\mu = \frac{e^{-\frac{J}{4g\sin\frac{p}{2}}}}{2g\sin\frac{p}{2}}
  \cdot \frac{\eta_1}{\tilde\eta_1}\frac{\eta_2}{\tilde\eta_2} \cdot \sigma(x_1,x_2).
\end{equation}

The value of the dressing factor at the pole is given by the same
expression as in $\superN=4$, namely \cite{Janik:2007wt}
\begin{equation}
  \sigma^2(x_1,x_2) = -\frac{16g^2}{e^2} e^{-ip} \sin^4\frac{p}{2}.
\end{equation}
Putting things together the $\mu$-term is
\begin{align}
  \delta\mathcal E_a^\mu &= \frac{2 i}{e} \sin\frac{p}{2} e^{-\frac{J}{8g\sin\frac{p}{2}}}, &\text{string frame}, \\
  \delta\mathcal E_a^\mu &= \frac{2 i}{e} \sin\frac{p}{2} e^{-\frac{J}{8g\sin\frac{p}{2}}} e^{-i\frac{p}{2}}, &\text{spin chain frame}.
\end{align}
The correction to the dispersion relation should be real. Taking the
real part of the above expressions we get
\begin{align}
  \delta\mathcal E &= 0, &&\text{string frame}, \\
  \delta\mathcal E &= \frac{2}{e} \sin^2\frac{p}{2} e^{-\frac{J}{8g\sin\frac{p}{2}}} 
                    = 2 \sin^2\frac{p}{2} e^{-\frac{E}{4g\sin\frac{p}{2}}}, &&\text{spin chain frame}.
\end{align}
We can now compare this result to the result of the algebraic curve
calculation. If we consider a fundamental magnon with $Q=1$ and let $n
= 0$ in \eqref{eq:disp-rel-corr-alg-c} we get exactly the above result
from the spin chain frame. Choosing $n = p/4\pi$ gives a vanishing
correction, like in the string frame.

\subsection{$\grSU(2) \times \grSU(2)$ giant magnon}

In order to calculate the corrections to a multi-magnon state we need
the generalized \Luscher formula of \citet{Hatsuda:2008na}%
\footnote{Essentially the same formula was independently given by
  \citet{Bajnok:2008bm}.}%
. The two-magnon $\mu$-term is given by
\begin{equation}
  \delta\mathcal E^\mu_{a_1 a_2} = 2 \sum_b (-1)^{F_b} \lrsbrk{1 - \frac{\mathcal E_{a_1}'(p_1)}{\mathcal E_b'(q_1^*)}} e^{-i q_1^* J} 
    \res_{q = \tilde q_1^*} S^{b a_1}_{b a_1}(q^1,p_1) S^{b a_2}_{b a_2}(q_1^*,p_2) .
\end{equation}
Since the two magnons are in different $\grSU(2)$ sectors, one of the
S-matrices will be of the type $S^{AA}$ or $S^{BB}$, while the other
will be of the type $S^{AB}$ or $S^{BA}$. Hence the full S-matrix
factor will be of the form
\begin{equation}
  S_0(q, p) \tilde S_0(q, p) \hat S^{1 b}_{1 b}(q, p) \hat S^{1 b}_{1 b}(q, p).
\end{equation}
But this is the exact same structure as for the $\grSU(2|2)^2$
S-matrix of $\superN=4$. Moreover, the full $\mu$-term now has the
form of the one magnon correction in $\superN=4$. Thus we can just use
the result of \citet{Janik:2007wt} and write
\begin{equation}
  \delta\mathcal E= \Re \lrsbrk{
    -32 g \sin^2\frac{p}{2} e^{-\frac{E}{4g\sin\frac{p}{2}}} \lrbrk{\frac{\eta_1 \eta_2}{\tilde\eta_1 \tilde\eta_2}}^2
    }.
\end{equation}
Again there are two choices for the phase factors $\eta$:
\begin{align}
  \delta\mathcal E &= -32 g \sin^3\frac{p}{2} e^{-\frac{E}{4g\sin\frac{p}{2}}} && \text{string frame}, \\
  \delta\mathcal E &= -32 g \sin^3\frac{p}{2} \cos(p) e^{-\frac{E}{4g\sin\frac{p}{2}}} && \text{spin chain frame}.
\end{align}

\section{Comparing the results}

The calculation of the finite size corrections to the two magnon
configuration in $\grSU(2) \times \grSU(2)$ which we considered,
closely follows the calculation of finite size corrections for a
single magnon in $\AdS_5 \times \Sphere^5$. In the string frame our final result was
\begin{align}
  E &= 8 g \sin\frac{p}{2} \lrbrk{1 - 4 \sin^2\frac{p}{2} e^{-\frac{E}{4g\sin\frac{p}{2}}}} \\
    &= 2\sqrt{2\lambda} \sin\frac{p}{2} \lrbrk{1 - \frac{4}{e^2} \sin^2\frac{p}{2} e^{-\frac{J}{\sqrt{2\lambda}\sin\frac{p}{2}}}}
\end{align}

As in that case we find perfect agreement between the results of the
finite gap and \Luscher calculations. Similar to the $\grSU(2)$ magnon
there is a correspondence between the choice of frame for the S-matrix
when calculating the \Luscher term, and the choice of branch, or
mode number, in the finite gap system.

\section{Conclusions}

In this paper we studied the finite size corrections for giant magnon
states in the $\grSU(2) \times \grSU(2)$ sector using the algebraic
curve as well as the \Luscher $\mu$-term. For the case of one
excitation in each $\grSU(2)$, with both excitations carrying the same
momenta, the resulting corrections perfectly match those of previous
calculations \cite{Grignani:2008te,Lee:2008ui,Ahn:2008hj}.  It is
encouraging that both the algebraic curve and the \Luscher term give
the same result as a direct string theory calculation.

The result for a single $\grSU(2)$ magnon is a bit harder to
interpret, since the result of the \Luscher term is not real.  In
itself this could be a sign that some contributions, such as those of
the bound states, are missing.  However, the real part of the result
perfectly matches the result from the algebraic curve. Moreover the
choice of the string frame versus spin-chain frame in the $\grSU(2|2)$
S-matrix corresponds to different choices of the mode number of the
curve.%
\footnote{Also for $\superN=4$ the choice of basis for the S-matrix in
  the \Luscher term corresponds to a choice of mode numbers for the
  algebraic curve. However, the \Luscher term is real in the string
  frame, so only this case has been generally considered.} 
The agreement between the two calculations give a good consistency
check between the algebraic curve \cite{Gromov:2008bz} and the
S-matrix proposed in \cite{Ahn:2008aa}.

The generic correction is proportional to the R-charge $Q$, and not to
$g$ as in $\superN=4$. Hence the classical correction vanishes for
fundamental magnons. From the algebraic curve perspective, it seems
like setting $Q=0$ forces the finite size magnon curve back to a curve
describing an infinite $J$ magnon. An explicit sigma model
construction of a single finite size $\grSU(2)$ magnon might lead to
an interpretation of this result.

The exceptional case is when we have two magnons with equal
momenta. The corrections are then enhanced to become finite. In both
the \Luscher and finite gap calculations this can be traced back to
the appearance of extra singularities.

\bigskip
\subsection*{Acknowledgments}
\bigskip

We would like to thank J.\@ Minahan and R.\@ Janik for their comments
on the manuscript. OOS would also like to thank V.\@ Giangreco Marotta
Puletti for many interesting discussions.

\appendix

\section{Notation}
\label{sec:notation}
The $\grSU(4)$ Dynkin labels $[p_1, q, p_2]$ are related to the
operator length $L$ and the excitation numbers $M_u$, $M_v$ and $M_r$
by
\begin{equation}
  [p_1, q, p_2] = [L - 2M_u + M_r, M_u + M_v - 2M_r, L - 2M_v + M_r].
\end{equation}
We assign the $\grSO(6) \cong \grSU(4)$ R-charges $J_1$, $J_2$ and $J_3$ as
\begin{align}
  J_1 &= q + \frac{p_2 + p_1}{2} = L - M_r, \\
  J_2 &= \frac{p_2 + p_1}{2}     = L + M_r - M_u - M_v, \\
  J_3 &= \frac{p_2 - p_1}{2}     = M_u - M_v,
\end{align}
We also introduce the charges
\begin{equation}
  J = J_1 + J_2 = 2L - M_u - M_v
  \quad \text{ and } \quad
  Q = J_1 - J_2 = M_u + M_v - 2M_r.
\end{equation}

\section{Properties of algebraic curve}\label{ap:prop-of-alg-curve}

This appendix summarize some properties of the quasi-momenta of the
algebraic curve for $\superN=6$ superconformal Chern-Simons.

\begin{itemize}
\item dependence of quasi-momenta
\begin{equation}
  \begin{pmatrix} q_1(x) \\ q_2(x) \\ q_3(x) \\ q_4(x) \\ q_5(x) \end{pmatrix}
  =
  -\begin{pmatrix} q_{10}(x) \\ q_9(x) \\ q_8(x) \\ q_7(x) \\ q_6(x) \end{pmatrix}
\end{equation}

\item condition for cuts
\begin{equation}
  q_i(x + i\epsilon) - q_j(x - i\epsilon) = 2\pi n_{ij}
\end{equation}

\item synchronization of poles at $x = \pm 1$
\begin{equation}
  \begin{pmatrix} q_1(x) \\ q_2(x) \\ q_3(x) \\ q_4(x) \\ q_5(x) \end{pmatrix}
  =
  -\begin{pmatrix} q_{10}(x) \\ q_9(x) \\ q_8(x) \\ q_7(x) \\ q_6(x) \end{pmatrix}
  =
  \frac{1}{2} \frac{1}{x \mp 1}
  \begin{pmatrix} \alpha_{\pm} \\ \alpha_{\pm} \\ \alpha_{\pm} \\ \alpha_{\pm} \\ 0 \end{pmatrix}
  + \dotsb
\end{equation}

\item inversion symmetry ($m \in \Integers$)
\begin{equation}
  \begin{pmatrix} q_1(1/x) \\ q_2(1/x) \\ q_3(1/x) \\ q_4(1/x) \\ q_5(1/x) \end{pmatrix}
  =
  \begin{pmatrix} 0 \\ 0 \\ \pi m \\ \pi m \\ 0 \end{pmatrix}
  +
  \begin{pmatrix} -q_2(x) \\ -q_1(x) \\ -q_4(x) \\ -q_3(x) \\ +q_5(x) \end{pmatrix}
  =
  \begin{pmatrix} 0 \\ 0 \\ \pi m \\ \pi m \\ 0 \end{pmatrix}
  +
  \begin{pmatrix} +q_9(x) \\ +q_{10}(x) \\ +q_7(x) \\ +q_8(x) \\ -q_6(x) \end{pmatrix}
\end{equation}

\item asymptotic behavior at $x \to \infty$
\begin{equation}
  \begin{pmatrix} q_1(x) \\ q_2(x) \\ q_3(x) \\ q_4(x) \\ q_5(x) \end{pmatrix}
  =
  \frac{1}{2gx}
  \lrbrk{\begin{array}{l} E + S \\ E - S \\ L - M_r \\ L + M_r - M_u - M_v \\ M_v - M_u \end{array}}
  =
  \frac{1}{2gx}
  \begin{pmatrix} E + S \\ E - S \\ \phantom{+}J_1 \\ \phantom{+}J_2 \\ -J_3 \end{pmatrix}
\end{equation}

\end{itemize}

\bibliographystyle{oos}
\bibliography{refs,chern-simons}

\end{document}